\documentclass{tlp}
\usepackage[english]{babel}
\usepackage[stdmacro,llncs]{macro}

\title{State Space Computation and Analysis of Time Petri Nets}

\author[G. Gardey, O.H. Roux and O.F. Roux]{GUILLAUME GARDEY and OLIVIER H. ROUX and OLIVIER  F.ROUX\\
IRCCyN (Institut de Recherche en Communication et Cybernétique de  Nantes), \\
	UMR CNRS 6597 \\
	Université de Nantes, \'Ecole Centrale de Nantes, \'Ecole des Mines de Nantes, CNRS \\
	1, rue de la Noë B.P. 92101 -- 44321 NANTES cedex 3 -- France \\
  \email{\{guillaume.gardey,olivier-h.roux,olivier.roux\}@irccyn.ec-nantes.fr}}

\begin{document}
  
  \maketitle

\begin{abstract}
  The theory of {\em Petri Nets} provides a general framework to specify the behaviors of real-time
  reactive systems and {\em Time Petri Nets} were introduced to take also temporal specifications
  into account. We present in this paper a forward zone-based algorithm to compute the state space
  of a bounded Time Petri Net: the method is different and more efficient than the classical State
  Class Graph. We prove the algorithm to be exact with respect to the reachability problem.
  Furthermore, we propose a translation of the computed state space into a Timed Automaton, proved
  to be timed bisimilar to the original Time Petri Net. As the method produce a single Timed
  Automaton, syntactical clocks reduction methods (\textsc{Daws} and \textsc{Yovine} for instance)
  may be applied to produce an automaton with fewer clocks. Then, our method allows to model-check
  \ttpn by the use of efficient Timed Automata tools.
\end{abstract}


\section{Introduction}

\subsection*{Framework}
The theory of \textrm{Petri Nets} provides a general framework to specify the behaviors of real-time
reactive systems and time extensions were introduced to take also temporal specifications into
account.  The two main time extensions of Petri Nets are Time Petri Nets (\tpn)~\cite{merlin-phd-74}
and Timed Petri Nets~\cite{ramchandani-phd-74}. While a transition can be fired within a given
interval for \tpn, in Timed Petri Nets, transitions are fired as soon as possible. There are also
numerous ways of representing time. 
\tpn are mainly divided in \ptpn, \atpn and \ttpn where a time interval is relative to places
(\ptpn), arcs (\atpn) or transitions (\ttpn). Finally, Time Stream Petri Nets~\cite{Diaz-lncs-94}
were introduced to model multimedia applications.

Concerning the timing analysis of these three models ((T,P,A)--\tpn), few studies have been realized
about model-checking.

Recent works~\cite{abdulla-ICATPN-01,defrutos-ICATPN-00} consider Timed Arc Petri Nets where each
token has a clock representing its ``age''. Using a backward exploration algorithm
\cite{abdulla-TCS-01,finkel-LATIN-98}, it is proved that the coverability and boundedness are
decidable for this class of Petri Nets. However, they assume a lazy (non-urgent) behavior of the
net: the firing of a transition may be delayed even if its clock's value becomes greater than its latest
firing time, disabling the transition.

In~\cite{rokicki-phd,rokicki-cav-94}, \textsc{Rokicki} considers an extension of labeled Petri Nets
called Orbitals Nets: each transition of the \tpn (safe \ptpn) is labeled with a set of events
(actions). The state space is built using a forward algorithm very similar to \textsc{Alur} and
\textsc{Dill} region based method. \textsc{Rokicki} finally uses partial order method to reduce time
and space requirements for verification purpose. The semantics used is not formally defined and
seems to differ from another commonly adopted proposed by \textsc{Khansa}~\cite{khansa-wodes-96} for
\ptpn.

In this paper, we consider \ttpn in which a transition can be fired within a time interval. For this
model, boundedness is undecidable and works report undecidability results, or decidability under the
assumption of boundedness of the \ttpn (as for reachability, decidability~\cite{popova-JIPC-91}).

\subsection*{Related Works}

\noindent
\paragraph{\bf State Space Computation of a \ttpn.} The main method to compute the state space of a \ttpn
is the State Class Graph~\cite{menasche-thesis-82,berthomieu-IEEE-91}.  A class $C$ of a \ttpn
is a pair $(M,D)$ where $M$ is a marking and $D$ a set of inequalities called the firing domain. The
variable $x_i$ of the firing domain represents the firing time of the enabled transition $t_i$
relatively to the time when the class $C$ was entered in and truncated to nonnegative times. The
State Class Graph preserves markings~\cite{bertho-TACAS-03} as well as traces and complete traces
but can only be used to check untimed reachability properties and is not accurate enough for
checking \emph{quantitative} real-time properties. An alternative approach has been proposed by
\textsc{Yoneda} \textit{et al.}~\cite{yoneda-IEICE-98} in the form of an extension of equivalence
classes (\textrm{atomic classes}) which allow \ctl model-checking.
\textsc{Lilius}~\cite{lilius-ENTCS-99} refined this approach so that it becomes possible to apply
partial order reduction techniques that have been developed for untimed systems. \textsc{Berthomieu}
and \textsc{Vernadat}~\cite{bertho-TACAS-03} propose an alternative construction of the graph of
atomic classes of \textsc{Yoneda} applicable to a larger class of nets. In~\cite{okawa-yoneda-97},
\textsc{Okawa} and \textsc{Yoneda} propose another method to perform \ctl model-checking on \ttpn,
they use a region based algorithm on safe \ttpn without $\infty$ as latest firing time. Their
algorithm is based on the one of~\cite{alur-TCS-94} and aims at computing a graph preserving
branching properties. Nevertheless, the algorithm used to construct the graph seems inefficient
(their algorithm do code regions) and no result can be exploited to compare with other  methods.

\noindent
\paragraph{\bf From \ttpn to \ta.} Several approaches aim at translating a Time Petri Net into a Timed
Automaton in order to use efficient existent tools on \ta. In~\cite{cortes-isss-00}, \textsc{Cortès}
\textit{et al.} propose to transform an extension of \ttpn into the composition of several \ta. Each
transition is translated into an automaton (not necessarily identical due to conflict problems) and
it is claimed that the composition captures the behavior of the \ttpn. In~\cite{avocs-04},
\textsc{Cassez} and \textsc{Roux} propose another structural approach: each transition is translated
into a \ta using the same pattern. The authors prove the two models are timed bisimilar.
In~\cite{sava-alla-01}, \textsc{Sava} and \textsc{Alla} compute the graph of reachable markings of a
\ttpn. The result is a \ta. However, they assume the \ttpn is bounded and does not include
$\infty$ as latest firing time. No proof is given of the timed bisimilarity between the two models.
In~\cite{lime-roux-pnpm-03}, \textsc{Lime} and \textsc{Roux} propose a method for building the State
Class Graph of a bounded \ttpn as a \ta. They prove the \ttpn to be timed bisimilar to the generated \ta.

Considering the translation of \ttpn into \ta, in order to study model's properties, raises the
problem of the model-checking feasibility of the resulting \ta. The model-checking complexity on \ta
is exponential in the number of clocks of the \ta. The proposed transformation in
\cite{avocs-04,cortes-isss-00} is to build as many \ta as the number of transitions of the
\ttpn. Consequently, there are as many clocks as in the initial \ttpn. It has also to be considered
that reduction method~\cite{daws-yovine-96} can not be applied to the resulting \ta: the parallel
composition has to be computed first. Nevertheless, the construction of \ta is straightforward and
linear in the number of transitions of the \ttpn. Concerning the method in \cite{lime-roux-pnpm-03},
the resulting \ta has a lower number of clocks. The method we propose produces an automaton with
more clocks than the previous method but its computation is faster.

Such translations show that \tctl and \ctl are decidable for bounded \ttpn and that developed algorithms on
\ta may be extended to \ttpn.

\subsection*{Contributions}

This paper is devoted to presenting an alternative approach to the state space construction of a
\ttpn.  The method is mainly based upon the region graph algorithm of \textsc{Alur} and
\textsc{Dill} on Timed Automaton. We propose to use a derived method using zones to compute the
state space of the \ttpn. The  algorithm is proved to be exact with respect to the
reachability problem and we propose to translate the state space it computes into a Timed Automaton,
bringing so the power of \ta model-cheking algorithms  to \ttpn.

We first recall the semantics of \ttpn and present a forward zone-based algorithm that computes the
state space of a \ttpn. Next, we present the labeling of the state space that produces a \ta we
proved to be timed bisimilar to the original \ttpn. We then compare our method to other used methods
on \ttpn and show its advantages. Finally, some applications are proposed.

\section{Time Petri Nets}
\label{sec:definitions}

\subsection{Definitions}

Time Petri Nets (\ttpn) are a time extension of classical Petri Nets. Informally, with each transition
of the Net is associated a clock and a time interval. The clock measures the time since the
transition has been enabled and the time interval is interpreted as a firing condition: the transition
may fire if the value of its clock belongs to the time interval.

Formally:
\begin{definition}[\ttpn]
  A Time Petri Net is a tuple $(P,T,\pre{(.)},\post{(.)},\mathbb{\alpha},\beta,M_0)$ defined by:
\begin{itemize}
\item $P=\{p_1,p_2,\ldots,p_m\}$ is a non-empty set of {\rm places},
\item $T=\{t_1,t_2,\ldots,t_n\}$ is a non-empty set of {\rm transitions},
\item $\pre{(.)} : T\to \bbbn^P$ is {\rm the backward incidence function},
\item $\post{(.)} : T\to \bbbn^P$ is {\rm the forward incidence function},
\item $M_0 \in \bbbn^P$ is {\rm the initial marking} of the Petri Net,
\item $\alpha : T \to \bbbq_{\geq 0}$ is the function giving {\rm the earliest firing times} of transitions,
\item $\beta : T \to \bbbq_{\geq 0}\cup \{ \infty \}$ is the function giving {\rm the latest firing
    times} of transitions.
\end{itemize}
\end{definition} 

A Petri Net marking $M$ is an element of $\bbbn^P$ such that for all $p \in P$, $M(p)$ is the number
of tokens in the place $p$.

A marking $M$ enables a transition $t$ if: $M \ge \pre{t_i}$. The set of transitions
enabled by a marking $M$ is \enabled{M}.

A transition $t_k$ is said to be \textit{newly} enabled by the firing of a transition $t_i$ if
$M-\pre{t_i}+\post{t_i}$ enables $t_k$ and $M-\pre{t_i}$ did not enable $t_k$. If $t_i$ remains
enabled after its firing then $t_i$ is newly enabled. The set of transitions newly enabled by a
transition $t_i$ for a marking $M$ is noted $\newlyenabled{M}{t_i}$.

$v \in (\bbbr_{\geq 0})^T$ is a valuation of the system. $v_i$ is the time elapsed since the
transition $t_i$ has been newly enabled.

The semantics of \ttpn is defined as a Timed Transition Systems (TTS).  Firing a transition is a
discrete transition of the TTS, waiting in a marking, the continuous transition.

\begin{definition}[Semantics of a \ttpn]
  The semantics of a \ttpn is defined by the Timed Transition System
  \mbox{$\mathcal{S}=(Q,q_0,\to)$}:
  \begin{itemize}
  \item $Q=\bbbn^P \times (\bbbr_{\geq 0})^T$
  \item $q_0=(M_0,\bar{0})$
  \item $\to \in Q \times (T \cup \bbbr_{\geq 0}) \times Q$ is the transition relation including a discrete
    transition and a continuous transition.
  \begin{itemize}
  \item[$\bullet$] The continuous transition is defined $\forall d \in \bbbr_{\geq 0}$ by:\\
    $ (M,v)\xrightarrow{e(d)}(M,v') \ {\rm iff} \ 
      \begin{cases}
        v' = v + d \\
        \forall k \in [1,n] \; M \geq ^\bullet \! t_k \Rightarrow v_k' \leq \beta(t_k)
      \end{cases}$
    
    \item[$\bullet$] The discrete transition is defined $\forall t_i \in T$ by: \\
      $ (M,v)\xrightarrow{t_i}(M',v') \; {\rm iff} \;
      \begin{cases}
        M \geq \pre{t_i} \\
        M' = M - \pre{t_i} + \post{t_i} \\
        \alpha(t_i) \leq v_i \leq \beta(t_i) \\
        \forall k \in [1,n] \; v_k'=
        \begin{cases}
          0 \ {\rm if} \ t_k \in \ \newlyenabled{M}{t_i} \\
          v_k \ {\rm otherwise}
        \end{cases}
      \end{cases}$
  \end{itemize}
\end{itemize}
\end{definition}

\subsection{The State Class Method}

The main method for computing the state space of a Time Petri Net is the State Class Method
introduced by \textsc{Berthomieu} and \textsc{Diaz} in~\cite{berthomieu-IEEE-91}.

\begin{definition}[State Class]
  A State Class $C$ of a \ttpn is a pair $(M,D)$ where $M$ is a marking and $D$ a set of
  inequalities called the firing domain. The variable $x_i$ of the firing domain represents the firing
  time of the enabled transition $t_i$ relatively to the time when the class $C$ was entered in.
\end{definition}

The State Class Graph is computed iteratively as follows:

\begin{definition}
  Given a class $C=(M,D)$ and a firable transition $t_j$, the successor class $C'=(M',D')$ by the
  firing of $t_j$ is obtained by:
  \begin{enumerate}
  \item Computing the new marking $M'=M-\pre{t_j} +\post{t_j}$.
  \item Making variable substitution in the domain: $\forall i \neq j, \; x_i \leftarrow x_i'+x_j$.
  \item Eliminating $x_j$ from the domain using for instance the Fourier-Motzkin method.
  \item Computing a canonical form of $D'$ using for instance the Floyd-Warshall algorithm.
  \end{enumerate}
\end{definition}

In the state class method, the domain associated with a class is relative to the time when the class
was entered in and as the transformation (time origin switching) is irreversible, absolute values of
clocks cannot be obtained easily. The produced graph is an abstraction of the state space for which
temporal information has been lost and generally, the graph has more states than the number of
markings of the \ttpn. Transitions between classes are no longer labeled with a firing constraint
but only with the name of the fired transition: the graph is a representation of the untimed
language of the \ttpn.

\subsection{Limitations of the State Class Method}

As a consequence of the State Class Graph construction, sophisticated temporal properties are not
easy to check.  Indeed, the domain associated with a marking is made of relative values of clocks and
the function to compute domains is not bijective. Consequently, domains can not easily be used to
verify properties involving constraints on clocks.

In order to get rid of these limitations, several works construct a different State Class Graph by
modifying the equivalence relation between classes. To our knowledge, proposed
methods~\cite{bertho-TACAS-03} depend on the property to check. Checking \ltl or \ctl properties
will lead to construct different State Class Graphs.

Another limitation of methods and proposed tools to check properties is the need to compute the
whole state space while only the reachability of a given marking is needed (e.g. for safety
properties). The graph is then analyzed by a model-checker. The use of \ttpn observers is even more
costly: actually, for each property to be checked, a new State Class Graph has to be built and the
observer can dramatically increase the size of the state space.

\medskip In the next section we will present another method to compute the state space of a bounded
\ttpn. It will be used in a later section to propose a Timed Automaton that is timed bisimilar to
the original \ttpn. As the graph has exactly as many nodes as the number of reachable markings of
the \ttpn, we obtain a compact representation of the state space which may be efficiently
model-checked using \ta tools.

\section{A Forward Algorithm to Compute the State Space of a Bounded \ttpn}
\label{sec:algorithm}

The method we propose in this paper is an adaptation, proved to be exact, of the region based method
for Timed Automaton~\cite{alur-TCS-94,rokicki-phd}. This algorithm starts from the initial state and
explores all possible evolutions of the \ttpn by firing transitions or by elapsing a certain amount
of time.

First, we define a \textit{zone} as a convex union of regions as defined by \textsc{Alur} and
\textsc{Dill}~\cite{alur-TCS-94}. For short, considering $n$ clocks, a zone is a convex subset of
${\left (\bbbr_{\geq 0} \right )}^n$.  A zone could be represented by a conjunction of constraints
on clocks pairs: $x_i - x_j \sim c$ where $\sim \in \{<,\leq,=,\geq,>\}$ and $c \in \bbbz$.

\subsection{Our Algorithm: One Iteration}

Given the initial marking and initial values of clocks (null vector), timing successors are
iteratively computed by letting time pass or by firing transitions.

Let $M_0$ be a marking and $Z_0$ a zone. The computation of the reachable markings from $M_0$
according to the zone $Z_0$ is done as follows:

\begin{itemize}
\item Compute the possible evolution of time (future): $\future{Z_0}$. This is obtained by setting
  all upper bounds of clocks to infinity.
\item Select only the possible valuations of clocks for which $M_0$ could exist, \ie valuations of
  clocks must not be greater than the latest firing time of enabled transitions :
  \begin{center}
    $ Z'_0 = \overrightarrow{Z_0} \cap \left \{ \bigwedge_i \left \{ x_i \leq \beta_i \;| \;t_i \in
        enabled \left (M_0\right ) \right \} \right \}$
\end{center}
So, $Z'_0$ is the maximal zone starting from $Z_0$ for which the marking $M_0$ is legal according to the \ttpn semantics.
\item Determine the firable transitions: $t_i$ is firable if $Z'_0 \cap \left \{x_i \geq \alpha_i
  \right \}$ is a non empty zone.
\item For each firable transition $t_i$ leading to a marking $M_{0i}$, compute the zone entering the
  new marking:
    \begin{center} 
      $Z_i = \left ( Z'_0 \cap \left \{x_i \geq \alpha_i \right \} \right ) \left [ X_e:=0\right ]$,
      where $X_e$ s the set of clocks of newly enabled transitions.
    \end{center}
    This means that each transition which is newly enabled has its clock reset. Then, $Z_i$ is a
    zone for which the new marking $M_{0i}$ is reachable.
\end{itemize}

\subsection{Convergence Criterion}
To ensure termination, a list of zones is associated with each reachable marking. It will keep track
of zones for which the marking was already analyzed or will be analyzed. At each step, we compare
the zone currently being analyzed to the ones previously computed. If the zone is included in one of
the list there is no need to go further because it has already been analyzed or it will lead to
compute a subgraph.

\subsection{Unboundedness in \ttpn}
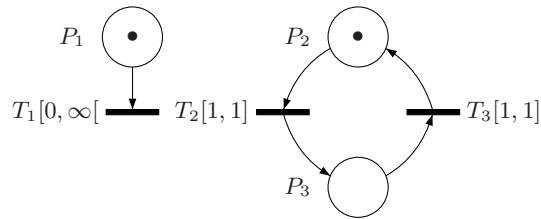
\begin{figure}[htbp]
  \centering
  \begin{picture}(30,30)(-15,-15)
    \node(P1)(-20,10){$\bullet$}
    \node(P2)(10,10){$\bullet$}
    \node(P3)(10,-10){}
    \gasset{NLdist=8}
    \nodelabel[NLangle=180](P1){$P_1$}
    \nodelabel[NLangle=180](P2){$P_2$}
    \nodelabel[NLangle=180](P3){$P_3$}
    
    \gasset{Nw=7,Nh=.7,Nmr=0,fillgray=0} 
    \gasset{ExtNL=y,NLdist=1,NLangle=180} 
    
    \node(T1)(-20,0){$T_1 [0,\infty[$}
    \node(T2)(0,0){$T_2 [1,1]$}
    \gasset{ExtNL=y,NLdist=1,NLangle=0} 
    \node(T3)(20,0){$T_3 [1,1]$}

    \drawedge(P1,T1){}
    \gasset{curvedepth=-2}
    \drawedge(P2,T2){}
    \drawedge(T2,P3){}
    \drawedge(P3,T3){}
    \drawedge(T3,P2){}
  \end{picture}
  \caption{Time Petri Net with an unbounded number of zones}
  \label{fig:exunb}
\end{figure}

An algorithm to enumerate reachable markings for a bounded \ttpn could be based on the described
algorithm but, generally, it will lead to a non-terminating computation. Though the number of
reachable markings is finite for a bounded \ttpn, the number of zones in which a marking is
reachable is not necessarily finite (see figure~\ref{fig:exunb}).

Let us consider the infinite firing sequence: $(T_2, T_3)^*$. The initial zone is $\{ x_1 = 0 \wedge
x_2 = 0 \wedge x_3 = 0 \}$ (where $x_i$ is the clock associated with $T_i$), the initial marking
$M_0=(P_1,P_2,P_3)=(1,1,0)$. By letting time pass, $M_0$ is reachable until $x_2 = 1$. When $x_2 =
x_1 = 1$ the transition $T_2$ has to be fired. The zone corresponding to clock values is: $Z_0 =
\{0 \leq x_1 \leq 1 \wedge x_1-x_2=0 \}$. By firing $T_2$ and then $T_3$, the net returns to its
initial marking. Entering it, values of clocks are: $x_1 = 2$, $x_2 = 0$ and $x_1-x_2=2$. Indeed,
$T_1$ remains enabled while $T_2$ and $T_3$ are fired and $x_2$ is reset when $T_3$ is fired because
$T_2$ is newly enabled. Given these new values, the initial marking can exists while $x_2 \leq
1$ \ie for the zone: $Z_1 = \{ 2 \leq x_1 \leq 3 \wedge x_1-x_2=2 \}$. By applying infinitely the
sequence $(T_2$, $T_3)$, there exists an infinite number of zones for which the initial marking is
reachable.

Actually, the number of zones is not bounded because infinity is used as latest firing time ($T_1$).
If infinity is not used as latest firing time, all clocks are bounded and so, the number of
different zones is bounded~\cite{alur-TCS-94}. The ``naive'' algorithm is then exact and can be used
to compute the state space of a bounded \ttpn.

\begin{consequence}
  For a bounded \ttpn without infinity as latest firing time, this forward analysis algorithm using
  zones computes the exact state space of the \ttpn.
\end{consequence}

In the next section, we propose a more general algorithm which computes the state space of a \ttpn
as defined in section~\ref{sec:definitions}, \ie with infinity as latest firing time allowed.

\subsection{General Algorithm}

A common operator on zones is the \textit{k-approx} operator. For a given $k$ value, the use of this
operator allows to create a finite set of distinct zones. The
algorithm proposed is an extension of the one presented in the previous section. It consists in
applying the \textit{k-approx} operator on the zone resulting from the last step:
\begin{equation*}
  Z_i = k-approx \left ( \left ( Z'_0 \cap \left \{x_i \geq \alpha_i \right \} \right ) \left [
  X_e:=0\right ] \right )
\end{equation*}

This approximation is based on the fact that once the clock associated with an unbounded transition
($[ \alpha, \infty [$) has reached the value $\alpha$, its precise value does not matter anymore.

Unfortunately recent works on Timed Automaton~\cite{bouyer-rr-02,bouyer-stacs-03} proved that this
operator generally leads to an overapproximation of the reachable localities of \ta.  However,
for a given class of \ta (diagonal-free), there is no overapproximation of the reachable localities.

Results of \textsc{Bouyer} are directly extensible for \ttpn. As computation on zones only involved
diagonal-free constraints, the following theorem holds:

\begin{theorem}
\label{th:main}
A forward analysis algorithm using \textit{k-approx} on zones is exact with respect to \ttpn marking
reachability for bounded \ttpn.
\end{theorem}

A detailed proof is available in~\cite{gardey-FORMATS-03}.

\subsection{Example} 
Let us consider the \ttpn of figure~\ref{fig:exunb}.

We associate the clock $x_i$ with the transition $T_i$ of the \ttpn and recall that clocks
associated with each transition count the time since the transition has been newly enabled.

The algorithm starts from the initial state: $l_0=(M_0,Z_0)$, with $M_0 = \begin{pmatrix} 1 &1 &0
\end{pmatrix}$ and $Z_0=\{ x_1 = x_2 = 0\}$. At marking $M_0$, transitions $T_1$ and $T_2$
are enabled.
  
The first step is to compute the possible future, \ie the maximal amount of time for which the
marking $M_0$ may exist:
\begin{eqnarray*}
  \overrightarrow{Z_0}  \cap Inv(M_0) & = & \left \{ x_1 = x_2 \in [0,\infty[ \right\} \cap
  \left\{ x_1 \leq \infty \wedge x_2 \leq 1 \right\} \\
  & = & \left \{ x_1 = x_2 \in [0,1] \right \}
\end{eqnarray*}

From this zone, two transitions are firable: $T_1$ and $T_2$.
\begin{description}
\item[Firing of $T_1$]\hfill
  \begin{itemize}
  \item the new marking is $M_1 = \begin{pmatrix} 0 &1 &0\end{pmatrix}$
  \item the new zone is obtained by intersecting the previous zone ($\future{Z_0} \cap Inv(M_0)$) with
    the guard $x_1 \geq 0$, deleting clocks of transitions that are no longer enabled in $M_1$
    ($x_1$) and reseting clocks of newly enabled transitions (none).
    \begin{eqnarray*}
      Z_1  =&  \left \{ x_1 = x_2 \in [0,1] \right \} \cap \left\{x_1\geq 0\right \} & (\text{intersect with guard}) \\
      = & \left \{ x_1 = x_2 \in [0,1] \right \}  & \\
      = & \left \{ x_2 \in [0,1] \right \}  & (\text{delete $x_1$})
    \end{eqnarray*}
  \end{itemize}
  
\item[Firing of $T_2$]\hfill
  \begin{itemize}
  \item the new marking is $M_3 = \begin{pmatrix} 1 &0 &1\end{pmatrix}$
  \item the new zone is obtained by intersecting the previous zone ($\future{Z_0} \cap Inv(M_0)$) with
    the guard $x_2 \geq 1$, deleting clocks of transitions that are no longer enabled in $M_3$
    ($x_2$) and reseting clocks of newly enabled transitions ($x_3$).
    \begin{eqnarray*}
      Z_3  =&  \left \{ x_1 = x_2 \in [0,1] \right \} \cap \left\{x_2\geq 1\right \} & (\text{intersect with guard}) \\
      = & \left \{ x_1 = x_2 = 1 \right \}  & \\
      = & \left \{ x_1 = 1 \right \}  & (\text{delete $x_2$}) \\
      = & \left \{ x_1 = 1 \wedge x_3 = 0 \right \} & (\text{reset $x_3$})
    \end{eqnarray*}
  \end{itemize}
\end{description}
We got two new states to analyze: $(M_1,Z_1)$ and $(M_3,Z_3)$. We apply the same algorithm to these
two states.
  
Considering $(M_1,Z_1)$:
\begin{eqnarray*}
  Z_1'=\overrightarrow{Z_1} \cap Inv(M_1) & = & \left \{ x_2 \in [0,1] \right \} \cap \{ x_2 \leq 1 \}\\
  & = & \left \{ x_2 \in [0, 1] \right \}
\end{eqnarray*}
$T_2$ is firable and leads to the new state: $(M_2,Z_2)$ with $M_2=\begin{pmatrix}0 &0
  &1\end{pmatrix}$ and $Z_2 = \{ x_3 = 0 \}$.  Analyzing $(M_2,Z_2)$ leads to the new state $(M_1,\{
x_2 = 0 \})$. As $\{x_2 = 0 \} \subset Z_1$, the algorithm stops and get a new state to analyze: $\left(M_3,Z_3\right)$.

Considering $(M_3,Z_3)$:
\begin{eqnarray*}
  Z_3'=\overrightarrow{Z_3} \cap Inv(M_3) & = & \left \{ x_1 - x_3 = 1,\; x_1 \in [0,\infty[ \right \} \cap \{ x_1
  \leq \infty \wedge x_3 \leq 1 \}\\
  & = & \left \{ x_1 -x_3 = 1 \wedge x_3 \leq 1 \right \}
\end{eqnarray*}
$T_3$ and $T_1$ are firable...

The analysis is performed until no new states are created. We then build the following graph of
reachable markings.

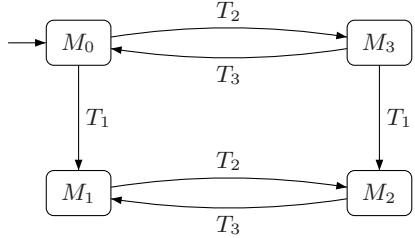
\begin{figure}[htbp]
  \centering
  \begin{picture}(40,30)(0,-15)
    \gasset{Nadjust=w,Nadjustdist=2,Nh=6,Nmr=1}
    \node[Nmarks=i](M0)(0,10){$M_0$}
    \node(M1)(0,-10){$M_1$}
    \node(M2)(40,-10){$M_2$}
    \node(M3)(40,10){$M_3$}

    \drawedge(M0,M1){$T_1$}
    \drawedge(M3,M2){$T_1$}
    \drawedge[curvedepth=2](M1,M2){$T_2$}
    \drawedge[curvedepth=2](M2,M1){$T_3$}
    \drawedge[curvedepth=2](M0,M3){$T_2$}
    \drawedge[curvedepth=2](M3,M0){$T_3$}
  \end{picture}
  \caption{Graph of reachable markings}
\end{figure}
\vspace{-1em}

\medskip In this section we have presented an algorithm that exactly computes the reachable markings
of a bounded \ttpn with $\infty$ as latest firing time. The graph computed is not suitable to verify
time logic properties. So, in the next section we present a transformation of the graph into a
Timed Automaton we proved to be timed bisimilar to the original \ttpn. Consequently, model-checking
methods on \ta become available for the model-checking of \ttpn.

\section{Marking Timed Automaton of Time Petri Net}

We first recall the definition of Timed Automata, introduced by \textsc{Alur} and
\textsc{Dill}~\cite{alur-TCS-94} and their semantics.

\subsection{Timed Automaton: Definitions}

Timed Automata are an extension of classical automata providing timing constraints. A transition can
occur if clocks valuations satisfy constraints called ``guard''.  Actions on clocks (reset for instance)
are associated with transition. The system can idle in a locality if valuations of clocks satisfy some
constraints called ``invariant''.

\begin{definition}[Constraints]
  Let $V$ be a set of clocks, $\mathcal{C}(V)$ is the set of timing constraints upon $V$ {\rm i.e.}
  the set of expressions $\delta$ defined by:
  \begin{eqnarray*}
    \delta:=v \sim c \; | \; v -v' \sim c \; | \; \neg \; \delta_1 \;
    | \; \delta_1 \wedge \delta_2  
  \end{eqnarray*}
  with $v,v' \in V$, $ \sim \in \{<,\leq,=,\geq,>\}$ and $c \in \bbbn$.
 \label{def:constraints}
\end{definition}

\begin{definition}[\ta]
  A Timed Automaton is a tuple $(L,l_0,C,A,E,Inv)$ defined by:
\begin{itemize}
\item $L$ a finite set of {\rm locations},
\item $l_0 \in L$ the {\rm initial location },
\item $C$ a finite set of positive real-valued {\rm clocks},
\item $A$ a finite set of {\rm actions},
\item $E \subset L \times \mathcal{C}(C) \times A \times 2^C \times L$ a finite set of {\rm
    transitions}. $e=(l,\gamma,a,R,l')$ is the transition from location $l$ to location $l'$ with
  the guard $\gamma$, the label $a$ and the set of clocks to reset $R$,
\item $Inv : L \times \mathcal{C}(C) \to \{true, false\}$, a function assigning to each location an
  {\rm invariant}.
\end{itemize}
\end{definition}

The semantics of a Timed Automaton is given by a Timed Transition System (TTS).

\begin{definition}[Semantics of a \ta]
  The semantics of a Timed Automaton is the Timed Transition System
  \mbox{$\mathcal{S}=(Q,Q_0,\rightarrow)$} where:
\begin{itemize}
\item $Q = L \times (\bbbr_{\geq 0})^C$,
\item $Q_0 = (l_0, \bar{0})$,
\item $\rightarrow$ is the transition relation including a discrete transition and a continuous
  transition.
  \begin{itemize}
  \item[$\bullet$] The discrete transition is defined $\forall a \in A$ by:
    \begin{equation*}
      (l,v) \xrightarrow{a} (l',v') \; {\rm iff} \; \exists
      (l,\gamma,a,R,l') \in E \; \text{such as :} \\
       \begin{cases} 
         \gamma (v) = true \\ 
         v' = v [ R \leftarrow 0 ] \\
         Inv(l')(v') = true
    \end{cases}
  \end{equation*}
  
\item[$\bullet$] The continuous transition is defined $\forall d \in \bbbr_{\geq 0}$ by:
    \begin{equation*}
    (l,v)  \xrightarrow{\epsilon(d)} (l,v') \; {\rm iff}
    \begin{cases}
      v'=v+d \\
      \forall t' \in [0,d], \; Inv(l)(v+t')=true
    \end{cases}
    \end{equation*}
    
  \end{itemize}
\end{itemize}
\end{definition}

\subsection{Labeling algorithm}

The algorithm given in section 3 represents the marking graph of the \ttpn. We
show here that it can easily be labeled to generate a Timed Automaton timed bisimilar to the \ttpn.

Let $\mathcal{G} = \left ( M, T \right )$ be the graph produced by the algorithm where:
\begin{itemize}
\item $M$ is the set of reachable markings of the \ttpn: $M_0, \dots, M_p$
\item $T$ is the set of transitions: $T_0, \dots, T_q$.
\end{itemize}

The Timed Automaton will be obtained by associating to each marking an invariant and to each
transition a guard and some clocks assignments.

\subsubsection{Invariant}
First, an invariant is associated with each marking $M_k$. By construction, in each marking, only the
possible evolution of time is computed: the entering zone is intersected with the set of
constraints $\left \{ x_i \leq \beta_i \right \}$, where $x_i$ are clocks of transitions enabled by
the marking $M_k$. Then, the invariant associated with each marking $M_k$ is defined by:
\begin{center}
  $I\leftp M_k \rightp = \left\{ x_i \leq \beta_i \; | \; t_i \in enabled\leftp M_k\rightp\right\}$
\end{center}

\subsubsection{Guard}

Each transition $T_k$ of the graph $\mathcal{G}$ corresponds to the firing of a transition $t_i$.
Then we label $T_k$ by:
\begin{itemize}
\item the action name $t_i$,
\item the guard: $x_i \geq \alpha_i$,
\item the clocks assignments: $x_k \leftarrow 0$ for all clocks $x_k$ associated with a newly enabled
  transition $t_k$
\end{itemize}

\subsection{Marking Timed Automaton}

The Timed Automaton we obtain is then defined as follows:

\begin{definition}[Marking Timed Automaton]
  \begin{itemize}
  \item $L = \left \{ M_0, \dots, M_p \right \}$ is the set of localities \ie the set of reachable
    markings of the \ttpn.
  \item $l_0 = M_0$ is the initial locality.
  \item $C = \left \{ x_1, \dots , x_q \right \}$ is the set of clocks \ie the set of all clocks
    associated with a transition.
  \item $A = \left \{ t_1, \dots, t_q \right \}$ is the set of actions \ie the transitions of the
    \ttpn.
  \item $E \subset L \times \mathcal{C}(C) \times A \times 2^C \times L$ is the finite set of
    transitions. Let $e= \left ( M_i,\gamma,a,R,M_j\right )$ a transition, $e$ is defined as follows:
    \begin{itemize}
    \item $a = t_k$
    \item $\gamma = x_k \geq \alpha_k$
    \item $R = \left \{ x_i \; | \; t_i \in \newlyenabled{M_i}{t_k} \right \}$
    \end{itemize}
  \item $Inv : L \times \mathcal{C}(C) \to \{true, false\}$, with:
    \begin{center}
      $Inv(M_i) = \left \{ x_i \leq \beta_i \; | \; t_i \in \enabled{M_i} \right \}$
    \end{center}
  \end{itemize}
\end{definition}

\paragraph{\bf Example}\hfill

Considering the \ttpn of figure~\ref{fig:exunb}, the resulting Timed Automaton is:

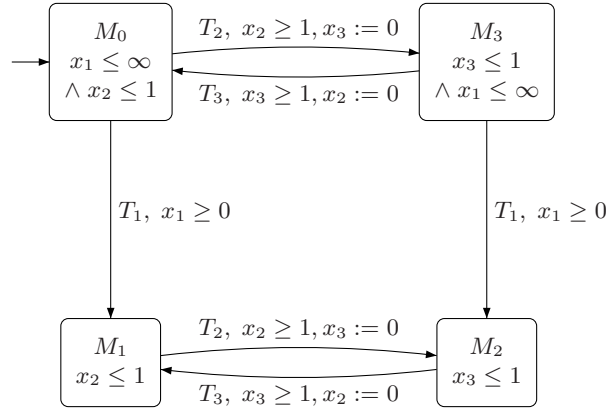
\begin{figure}[htbp]
  \centering
  \begin{picture}(50,50)(0,-25)
    \gasset{Nadjust=wh,Nadjustdist=2,Nh=6,Nmr=1}
    \node[Nmarks=i](M0)(0,20){$\begin{matrix}M_0\\ x_1 \leq \infty \\\wedge \;x_2 \leq 1  \end{matrix}$}
    \node(M1)(0,-20){$\begin{matrix}M_1\\x_2\leq 1\end{matrix}$}
    \node(M2)(50,-20){$\begin{matrix}M_2\\x_3\leq 1 \end{matrix}$}
    \node(M3)(50,20){$\begin{matrix}M_3\\x_3\leq 1 \\ \wedge \; x_1 \leq \infty\end{matrix}$}

    \drawedge(M0,M1){$T_1, \; x_1 \geq 0$}
    \drawedge(M3,M2){$T_1, \; x_1 \geq 0$}
    \drawedge[curvedepth=2](M1,M2){$T_2, \; x_2 \geq 1, x_3 := 0$}
    \drawedge[curvedepth=2](M2,M1){$T_3, \; x_3 \geq 1, x_2 := 0$}
    \drawedge[curvedepth=2](M0,M3){$T_2, \; x_2 \geq 1, x_3 := 0$}
    \drawedge[curvedepth=2](M3,M0){$T_3, \; x_3 \geq 1, x_2 := 0$}
  \end{picture}
  \caption{Time Marking Automaton}
\end{figure}

\subsection{Bisimulation}

\begin{definition}
  As defined in the time transition system for a \ttpn $\mathcal{T}$, we note
  $\mathcal{Q}_{\mathcal{T}}$ the set of states of $\mathcal{T}$. $\mathcal{Q}_{\mathcal{A}}$ is the
  set of states of a \ta $\mathcal{A}$.
\end{definition}

\begin{definition}
  Let $\mathcal{R} \subset \mathcal{Q}_{\mathcal{T}} \times \mathcal{Q}_{\mathcal{A}}$ be the
  relation between a state of the Timed Automaton and a state of the Time Petri Net defined by:
\begin{equation*}
  \begin{cases}
    \forall (M,v) \in \mathcal{Q}_{\mathcal{T}} \\
    \forall (l,\bar{v}) \in \mathcal{Q}_{\mathcal{A}}
  \end{cases}, \;
  (M,v) \mathcal{R} (l,\bar{v})  \Leftrightarrow 
  \begin{cases}
    M = \textbf{M}(l) \\
    v = \bar{v}
  \end{cases}
\end{equation*}
where \textbf{M} is the function giving the associated marking of a \ta state $l$.
\end{definition}

Two states are in relation if their ``markings'' and their clocks valuations are equals.

\begin{theorem}
  $\mathcal{R}$ is a bisimulation:
  
  For all $\left(M,v\right)$, $\left(l,\bar{v}\right)$ such that $\left(M,v\right) \mathcal{R}
  \left(l,\bar{v}\right)$:
  \begin{itemize}
  \item $(M,v) \xrightarrow{t_i} (M',v')$ $\Leftrightarrow$
    $\begin{cases}
      (l,\bar{v})\xrightarrow{t_i}(l',\bar{v}')\\
      (M',v')\mathcal{R}(l',\bar{v}')
      \end{cases}$
  \item $(M,v) \xrightarrow{\delta} (M,v')$ $\Leftrightarrow$
    $\begin{cases}
      (l,\bar{v})\xrightarrow{\delta}(l,\bar{v}')\\
      (M,v')\mathcal{R}(l,\bar{v}')
      \end{cases}$
  \end{itemize}
  
\end{theorem}

\begin{proof}
  
  {\it \underline{Continuous transition -- time elapsing}.}
 
  Let $(M,v_{\mathcal{T}}) \in \mathcal{Q}_{\mathcal{T}}$, $(l,v_{\mathcal{A}}) \in
  \mathcal{Q}_{\mathcal{A}}$, and $\delta \in \bbbr^{\geq 0}$.
 
  \medskip We prove that if the \ttpn can idle in a state, this is allowed on the constructed \ta
  \textit{i.e.}  if the system can idle for any $\delta$ such that $\forall k \in [1,n] \; M \geq
  {}^\bullet t_k \Rightarrow v_{\mathcal{T}}(t_k)+\delta \leq \beta(t_k)$ then the automaton
  verifies: $\forall t \in [0,\delta] \; Inv(l)(v_{\mathcal{A}}+t)=true$.
  
  By construction, the invariant of the location $l$ is obtained by the conjunction of the latest
  firing times of enabled transitions. So $Inv(l)= \bigwedge \left \{ x_i \leq \beta (t_i) \right
  \}$ where $t_i \in enabled(\textbf{M}(l))$. $(M,v_{\mathcal{T}})$ and $(l,v_{\mathcal{A}})$ are in
  relation so $v_{\mathcal{T}}=v_{\mathcal{A}}$. As $v_{\mathcal{T}}(t_i) + \delta \leq \beta(t_i)$
  then for all $t \in [0,\delta] \; v_{\mathcal{A}}(t_i) + t \leq \beta (t_i)$. This means that
  $\forall t \in [0,\delta] \; Inv(l)(v_{\mathcal{A}}+t)=true$.
  
  To conclude, the automaton can idle in the state and $(M,v_{\mathcal{T}}+\delta) \mathcal{R}
  (l,v_{\mathcal{A}}+\delta)$.
  
  \medskip {\it Symmetrically}, we prove that if the \ta can idle for a time $\delta$, the \ttpn can
  idle for the same time $\delta$.
  
  According to the semantics of \ttpn, a continuous transition can occur if and only if $\forall t_k
  \in enabled(M), \; v_{\mathcal{T}}(t_k)+\delta \leq \beta(t_k)$. As $(M,v_{\mathcal{T}})$ and
  $(l,v_{\mathcal{A}})$ are in relation, $v_{\mathcal{T}}=v_{\mathcal{A}}$. The \ta can idle in the
  state for all $t \in [0,\delta] \; v_{\mathcal{A}}(t_i) + t \leq \beta (t_i)$ by construction of
  the invariant. Then, $t=\delta$ prove the result.
  
  The \ttpn can idle in the marking and $(M,v_{\mathcal{T}}+\delta) \mathcal{R}
  (l,v_{\mathcal{A}}+\delta)$.
  
  \medskip Concerning continuous transitions, $\mathcal{R}$ is a bisimulation.
  
  \medskip
  
  {\it \underline{Discrete transition -- firing a transition $t_i$}} Let $(M,v_{\mathcal{T}}) \in
  \mathcal{Q}_{\mathcal{T}}$ and $(l,v_{\mathcal{A}}) \in \mathcal{Q}_{\mathcal{A}}$ be two states
  in relation.
  
  \medskip We prove that if a transition is firable for the \ttpn, it is firable for the \ta and the
  two resulting states are in relation.
  
  A transition $t_i$ of the \ttpn can be fired if: $M \geq {}^\bullet t_i$ and $\alpha(t_i) \leq
  v_{\mathcal{T}}(t_i) \leq \beta(t_i)$. The resulting marking is $M'=M-{}^\bullet t_i+t_i^\bullet$
  and the resulting valuation is $v'_{\mathcal{T}}(t_k)=0$ for all newly enabled transition $t_{k}$,
  all others valuations remain unchanged.
  
  The corresponding action is allowed on the constructed \ta if and only if
  \begin{equation*}
    \exists (l,\gamma,a,R,l') \in E \; \text{such as :}
    \begin{cases}
      \gamma (v) = true  \\
      v_{\mathcal{A}} = v_{\mathcal{A}} [ R \leftarrow 0 ]  \\
      Inv(l')(v'_{\mathcal{A}}) = true
    \end{cases}
  \end{equation*}
  
  As $t_i$ is firable, it exits by construction a transition of the \ta from $l$, such that
  $\textbf{M}(l)=M$, to a location $l'$ such that $\textbf{M}(l')=M'$. The guard is by construction,
  $\gamma = x_i \geq \alpha(t_i)$. Thus, as $t_i$ is firable $\gamma(v_{\mathcal{A}})=true$.
  
  Also by construction, the clocks to be reset for the \ta are the same clocks to be reset for the
  \ttpn. Thus, $v'_{\mathcal{A}}=v'_{\mathcal{T}}$.
 
  As clocks newly enabled are set to 0, they verifies the inequalities $x_j \leq \beta(t_j)$ in the
  invariant of $l'$. All other clocks stay unchanged: $v'_{\mathcal{A}}(t_j) \leq \beta(t_j)$ for
  all other enabled clocks. Thus, $Inv(l')(v'_{\mathcal{A}})=true$.
  
  So the transition on \ta is allowed and $(M',v'_{\mathcal{T}})\mathcal{R}(l',v'_{\mathcal{A}})$.
  
  \medskip {\it Symmetrically}, we prove that if $t_i$ is firable for the \ta, it is firable for the
  \ttpn. The two resulting states are in relation.
  
  A transition $e=(l,t_i,\gamma,R,l')$ of the \ta can occur and leads to a new state
  $(l',v'_{\mathcal{A}})$ if and only if $\gamma (v_{\mathcal{A}}) = true$ and
  $Inv(l')(v'_{\mathcal{A}}) = true$. Then $v'_{\mathcal{A}}=v_{\mathcal{A}}[R\leftarrow 0]$.
  
  The corresponding action is allowed on the \ttpn and leads to a new state $(M',v'_{\mathcal{T}})$
  if and only if:
  \begin{equation*}
    \begin{cases}
      M \geq ^\bullet \! t_i \\
      M' = M - ^\bullet \! t_i + t_i^\bullet \\
      \alpha(t_i) \leq v_i \leq \beta(t_i) \\
      \forall \; \text{transitions} \; t_k \; v'_{\mathcal{T}}(t_k)=
      \begin{cases}
        0 \ {\rm if} \ t_k \in \ \uparrow \! enabled(M,t_i) \\
        v_{\mathcal{T}}(t_k) \ {\rm otherwise}
      \end{cases}
    \end{cases}
  \end{equation*}
  
  By definition of the Marking Timed Automaton, if $t_i$ is firable for the \ta, it is for the
  \ttpn.  So $M \geq ^\bullet \! t_i$ and the resulting marking is by definition $M' = M - ^\bullet
  \!  t_i + t_i^\bullet$.
  
  $(l,v_{\mathcal{A}})$ and $(M,v_{\mathcal{T}})$ are in relation so
  $v_{\mathcal{T}}=v_{\mathcal{A}}$.
  
  As, $\gamma(v_{\mathcal{A}})=true$ and $Inv(l)(v_{\mathcal{A}})=true$ so, $\alpha(t_i) \leq
  v_{\mathcal{T}}(t_i) \leq \beta(t_i)$.
  
  By construction, the clocks to be reset are the clocks of newly enabled transitions \ie the clocks
  of $R$. So $v'_{\mathcal{A}}=v'_{\mathcal{T}}$.
  
  To conclude, $t_i$ is firable for the \ttpn and $(M',v'_{\mathcal{T}})$ and
  $(l',v'_{\mathcal{A}})$ are in relation.
  
  \medskip $\mathcal{R}$ is a bisimulation for discrete transitions.
\end{proof}

\section{Performances}

We have implemented the algorithm to compute all the reachable markings of a bounded \ttpn using
\dbm (Difference Bounded Matrices) to encode zones. The tool implemented (\mercutio) is integrated
into \romeo~\cite{romeo}, a software for \ttpn edition and analysis.

As boundedness of \ttpn is undecidable, \mercutio offers stopping criteria: number of reached
markings, computation time, bound on the number of tokens in a place. It also provides an on-the-fly
reachability test of markings and export the automaton in \kronos or \uppaal syntax. Concerning the on-the-fly 
reachability test, \mercutio also provides a trace (sequence of transitions and interval in
which they are fired) leading to the marking.

\subsection{Comparison with other methods}

We present here a comparison (Table~\ref{tab:comparison}) of three methods to compute the state
space of a \ttpn:

\begin{itemize}
\item the method proposed in this paper with our tool \mercutio.
\item the State Class Graph computation (\textsc{Berthomieu}) with the tool \tina.
\item the State Class Timed Automaton (\textsc{Lime} and \textsc{Roux}) with the tool \gpn.
\end{itemize}

\begin{table}
  \caption{Time to compute the state space of a \ttpn}
  \begin{tabular}{lcccc}\hline\hline
    Time Petri Net           & \ttpn (p./t.) & \tina       & \gpn        & \mercutio \\ \hline
    Example 1 (oex15)        & 16 / 16      & 10.5 s      &  12.9 s     & 2 s \\
    Example 2 (oex7)         & 22 / 20      & 30.5 s      &  9.8 s      & 1.3 s\\
    Example 3 (oex8)         & 31 / 21      & 29 s        &  12.2 s     & 1.4 s \\
    Example 4 (P6C7)         & 21 / 20      & 31.6 s      &  1 min 17 s & 7.9 s\\ 
    Example 5 (P10C10)       & 32 / 31      & 4.2 s       &  6.8 s      & 1 s\\ 
    Example 6 (GC - 3)       & 20 / 23      & 2 s         &  1.2 s      & 0.1 s\\ 
    Example 7 (GC - 4)       & 24 / 29      & 3 min 8 s   &  1 min 3 s  & 10.8 s\\ 
    Example 8 (P6C9)         & 25 / 24      & 2 min 49 s  &  6 min 2 s  & 22.9 s\\
    Example 9 (P6C10)       & 27 / 26       &  8 min 53 s  &  36 min      & 1 min\\
    Example 10 (P6C11)       & 29 / 28      &  14 min 36 s  & 1 h 1 min   & 2 min 20s\\
    Example 11 (P6C12)       & 31 / 30      &  23 min 34 s   & 2 h 7 min   & 3 min 59s \\
    Example 12 (P6C13)       & 33 / 32     &   36 min 25 s   & $\times$ & 6 min 3s \\
    \hline \hline
  \end{tabular}
  \label{tab:comparison}
\vspace{-1em}
\end{table}

Computations were performed on a Pentium 2 (400MHz) with 320MB of RAM.

Examples 1 to 5 come from real-time systems (parallel tasks [1], periodic tasks[2--3],
producer-consumer [4--5,8--12]). Examples 7 and 8 are the classical level crossing example (3
and 4 trains).

For this set of examples and for all nets we have tested, our tool performs better than \tina and
than \gpn. For example 12, \gpn ran out of memory.

\subsection{Reducing the number of clocks}

A major issue in model checking \ta is the number of clocks in the automaton. Time computation is
exponential in the number of clocks. Consequently, obtaining an automaton with a reduced number of
clocks is of importance.

The algorithm we propose assigns a clock to each transition. Thus, the resulting automaton has as
many clocks as transitions of the \ttpn. However we have underlined that for each location,
only a reduced number of clocks (active clocks) really matter for the timing evolution of the \ttpn.

\textsc{Daws} and \textsc{Yovine} in~\cite{daws-yovine-96} proposed a syntactical method to reduce
the number of clocks of a \ta. As a single Timed Automaton is build with our method (no need to compute
parallel composition) we applied this reduction. The table~\ref{tab:exoptikron} presents the
comparison between the clocks of (1) the Timed Automaton obtained, (2) the Timed Automaton obtained
after syntactical clocks reduction (we used \optikron from \kronos~\cite{kronos}), (3) the State
Class Timed Automaton using \gpn that ensures a minimal number of clocks using classes.

\begin{table}
  \caption{Structure of resulting Timed Automata}
  \begin{tabular}{lccccccc}
    \hline \hline
    \multirow{2}{30mm}{Time Petri Net} & \multirow{2}{20mm}{Clocks(1)\footnotemark} &
    \multicolumn{3}{c}{Marking(2) \ta} & \multicolumn{3}{c}{State Class \ta(3)} \\
     & & Cl. \footnotemark & N.\footnotemark & T.\footnotemark & Cl. & N. & T. \\ \hline
     Example 1 (oex15)        &  16                    & 4 & 361 & 1095         & 4  & 998 & 3086\\
     Example 2 (oex7)         &  20                    & 11 & 637 & 2284        & 7 & 1140 & 3990\\
     Example 3 (oex8)         &  21                    & 11 & 695 & 2444        & 7 & 1277 & 4344\\
     Example 4 (P6C7)         &  20                    & 13 & 449 & 4175        & 3 & 11490 & 50268\\ 
     Example 5 (P10C10)       &  31                    & 4  & 1088 & 5245       & 2 & 1088 & 5245\\ 
     Example 6 (GC - 3)       &  23                    & 5  & 94 & 271          & 3 & 286 & 763\\ 
     Example 7 (GC - 4)       &  29                    & 6  & 318 & 1221        & 4 & 2994 & 11806\\
     Example 8 (P6C9)         &  24                    & 15  & 1299 & 12674      & 3  & 24483 &
    117918\\
    Example 9 (P6C10)        &  26                  & 16 & 2596 & 27336 & 3 & 59756 & 313729\\
    Example 10 (P6C11)        & 28 & 17 & 4268&44620 & 3 & 82583 & 440540\\
    Example 11 (P6C12)        & 30 & 18 & 6846 & 70856 & 3 & 112023 & 606771\\ 
    Example 12 (P6C13)        & 32 & 19 & 10646 & 108842 & $\times$ &$\times$  & $\times$\\
    \hline \hline
   \end{tabular}
   \label{tab:exoptikron}
   \addtocounter{footnote}{-4}
   \raggedright Number of: 
   \tableFootnote{clocks of the original \ttpn}, \tableFootnote{clocks of the \ta},
   \tableFootnote{nodes of the \ta}, \tableFootnote{transitions of the \ta}.
\vspace{-1em}
\end{table}

These results are all the more encouraging that, reducing the number of clocks is made syntactically
and is made at no cost comparatively to the state space computation. The State Class Timed Automaton
always as a lower number of clocks but its construction is not as fast as our method: the Timed
Automaton has lower clocks at the price of a greater size. For example 12, we have not succeeded in
computing the State Class Timed Automaton (out of memory).

\section{Applications}

We propose in this section some applications of our method to model-check \ttpn.

\subsection{Model checking of Quantitative Properties}

Since they were introduced, Timed Automata are an active research area and several methods and tools
have been developed to analyze them. Tools like \uppaal~\cite{uppaal} or \kronos~\cite{kronos}
successfully implement efficient algorithms and data structures to provide model-checking on \ta
(\tctl model-checking for instance): numerous case studies have been performed with real reactive
systems.

Concerning \ttpn, few studies were realized and properties that can be checked are mainly
safety untimed properties (reachability).  Time or untime properties are mainly verified over \ttpn
using ``observers''. Basically, properties are transformed in an additional \ttpn motif called
``observer'', and then, the problem is transformed into a reachability test.  Such methods are not
easy to use: (1) modeling the property with an observer is not easy (it exists some generic
observers~\cite{toussaint-FTDCS-97}, but for few properties), (2) the observer's size may be as
large as the initial \ttpn, (3) due to the increase of the \ttpn's size, computing the state space
will be more time expensive.

The method we propose here, is to use existent \ta tools to perform model-checking of \ttpn. As a
Timed Automaton is produced, model-check a \ttpn (\ltl,\ctl) becomes possible and
verifying quantitative time property (\tctl) is possible. Moreover, as the automaton
constructed is a Timed Automaton with diagonal free constraints, model checking could be
done using on-the-fly algorithms on \ta (\textsc{Uppaal}\cite{uppaal}, \textsc{Kronos}\cite{kronos}).

\paragraph{\bf Example}\hfill

\begin{figure}
  \centering \subfigure[Controller]{
    \begin{picture}(60,60)(5,-60)
      \node(in)(0,-30){}
      \node(far)(40,-30){$n$}
      \node(coming)(40,-10){}
      \node(leaving)(40,-50){}
      \nodelabel[NLangle=180,NLdist=8](in){$in$}
      \nodelabel[NLangle=0,NLdist=8](far){$far$}
      \nodelabel[NLangle=90,NLdist=8](coming){$Coming$}
      \nodelabel[NLangle=-90,NLdist=8](leaving){$Leaving$}
      
      \node[Nw=0.7,Nh=7,Nmr=0,fillgray=0](app1)(20,-10){}
      \node[Nw=0.7,Nh=7,Nmr=0,fillgray=0](app2)(20,-23){}
      \node[Nw=0.7,Nh=7,Nmr=0,fillgray=0](exit1)(20,-37){}
      \node[Nw=0.7,Nh=7,Nmr=0,fillgray=0](exit2)(20,-50){}
      \node[Nw=0.7,Nh=7,Nmr=0,fillgray=0](down1)(60,-10){}
      \node[Nw=0.7,Nh=7,Nmr=0,fillgray=0](up1)(60,-50){}
      \nodelabel[NLangle=90,ExtNL=y,NLdist=1](app1){$App$}
      \nodelabel[NLangle=90,ExtNL=y,NLdist=1](app2){$App$}
      \nodelabel[NLangle=90,ExtNL=y,NLdist=1](exit1){$Exit$}
      \nodelabel[NLangle=90,ExtNL=y,NLdist=1](exit2){$Exit$}
      \nodelabel[NLangle=90,ExtNL=y,NLdist=1](up1){$Up\; [0,0]$}
      \nodelabel[NLangle=90,ExtNL=y,NLdist=1](down1){$Down\; [0,0]$}
      
      \drawedge(app1,coming){}
      \drawedge(coming,down1){}
      \drawedge(app1,in){}
      \drawedge(in,exit2){}
      \drawedge(exit2,leaving){}
      \drawedge(leaving,up1){}
      \drawedge(exit1,far){}
      \drawedge(far,app2){}
      \drawedge[curvedepth=2,ExtNL=n,ELpos=60,ELdistC=y,ELdist=1](far,app1){\scriptsize\colorbox{white}{$n$}}
      \drawedge[curvedepth=2,ExtNL=n,ELpos=60,ELdistC=y,ELdist=1](app1,far){\scriptsize\colorbox{white}{$n-1$}}
      \drawedge[curvedepth=2,ExtNL=n,ELpos=60,ELdistC=y,ELdist=1](far,exit2){\scriptsize\colorbox{white}{$n-1$}}
      \drawedge[curvedepth=2,ExtNL=n,ELpos=60,ELdistC=y,ELdist=1](exit2,far){\scriptsize\colorbox{white}{$n$}}
      \drawedge[curvedepth=2](in,app2){}
      \drawedge[curvedepth=2,ExtNL=n,ELpos=50,ELdistC=y,ELdist=0](app2,in){\scriptsize\colorbox{white}{$2$}}
      \drawedge[curvedepth=2](exit1,in){}
      \drawedge[curvedepth=2,ExtNL=n,ELpos=50,ELdistC=y,ELdist=0](in,exit1){\scriptsize\colorbox{white}{$2$}}
    \end{picture}
    \label{fig:controller}
  }\goodgap \subfigure[Barrier model]{
    \begin{picture}(50,60)(-5,-60)
      \node(open)(20,-5){$\bullet$}
      \node(lowering)(0,-30){}
      \node(raising)(40,-30){}
      \node(closed)(20,-55){}
      \nodelabel[NLangle=90,NLdist=8](open){$Open$}
      \nodelabel[NLangle=90,NLdist=8](lowering){$Lowering$}
      \nodelabel[NLangle=45,NLdist=9](raising){$Raising$}
      \nodelabel[NLangle=90,NLdist=8](closed){$Closed$}

      \node[Nw=7,Nh=0.7,Nmr=0,fillgray=0](down1)(0,-15){}
      \node[Nw=7,Nh=0.7,Nmr=0,fillgray=0](r)(40,-15){}
      \node[Nw=7,Nh=0.7,Nmr=0,fillgray=0](l)(0,-45){}
      \node[Nw=7,Nh=0.7,Nmr=0,fillgray=0](up)(40,-45){}
      \node[Nw=0.7,Nh=7,Nmr=0,fillgray=0](down2)(20,-30){}
      \nodelabel[NLangle=0,ExtNL=y,NLdist=1](l){$L \; [1,2]$}
      \nodelabel[NLangle=0,ExtNL=y,NLdist=1](r){$R \; [1,2]$}
      \nodelabel[NLangle=0,ExtNL=y,NLdist=1](up){$Up$}
      \nodelabel[NLangle=0,ExtNL=y,NLdist=1](down1){$Down$}
      \nodelabel[NLangle=90,ExtNL=y,NLdist=1](down2){$Down$}
      
      \drawedge(open,down1){}
      \drawedge(down1,lowering){}
      \drawedge(lowering,l){}
      \drawedge(l,closed){}
      \drawedge(closed,up){}
      \drawedge(up,raising){}
      \drawedge(raising,r){}
      \drawedge(r,open){}
      \drawedge(raising,down2){}
      \drawedge(down2,lowering){}
    \end{picture}
    \label{fig:barrier}
  }\\
  \subfigure[Train model]{
    \begin{picture}(45,35)(-5,-15)
      \node(closei)(-5,0){}
      \node(oni)(20,-10){}
      \node(lefti)(45,0){}
      \node(fari)(20,10){$\bullet$}
      \nodelabel[NLangle=180,NLdist=9](closei){$Close_i$}
      \nodelabel[NLangle=-90,NLdist=8](oni){$On_i$}
      \nodelabel[NLangle=0,NLdist=8](lefti){$Left_i$}
      \nodelabel[NLangle=90,NLdist=8](fari){$Far_i$}
      
      \node[Nw=0.7,Nh=7,Nmr=0,fillgray=0](ini)(5,-10){}
      \node[Nw=0.7,Nh=7,Nmr=0,fillgray=0](exi)(35,-10){}
      \node[Nw=0.7,Nh=7,Nmr=0,fillgray=0](exit)(35,10){}
      \node[Nw=0.7,Nh=7,Nmr=0,fillgray=0](app)(5,10){}
      \nodelabel[NLangle=-90,ExtNL=y,NLdist=1](ini){$In_i \; [3,5]$}
      \nodelabel[NLangle=-90,ExtNL=y,NLdist=1](exi){$Ex_i \; [2,4]$}
      \nodelabel[NLangle=90,ExtNL=y,NLdist=1](exit){$Exit \; [0,0]$}
      \nodelabel[NLangle=90,ExtNL=y,NLdist=1](app){$App$}

      \drawedge(closei,ini){}
      \drawedge(ini,oni){}
      \drawedge(oni,exi){}
      \drawedge(exi,lefti){}
      \drawedge(lefti,exit){}
      \drawedge(exit,fari){}
      \drawedge(fari,app){}
      \drawedge(app,closei){}
    \end{picture}
    \label{fig:traini}
  }
  \caption{Gate Controller}
  \label{fig:train}
\end{figure}
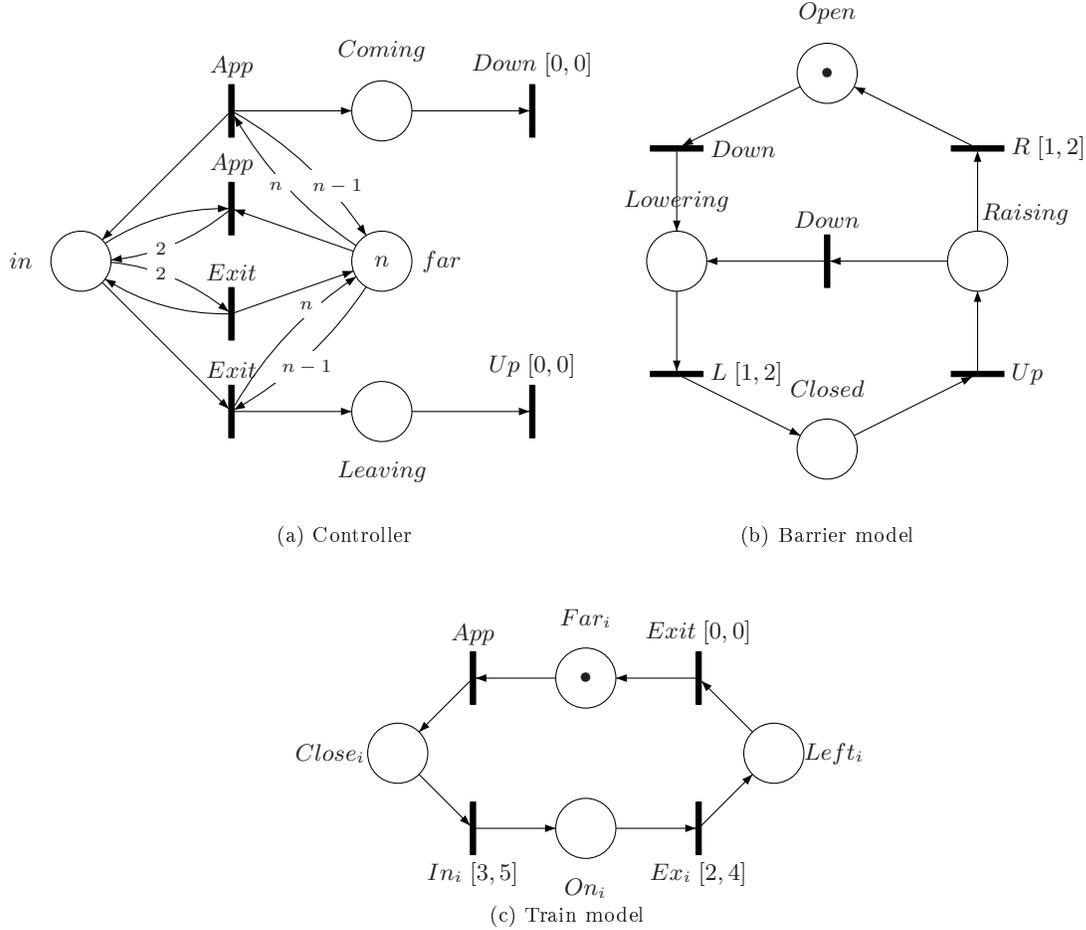

Let us consider the classical level crossing example. The system is modeled using the three patterns
of the figure~\ref{fig:train}. This model is made of a controller (\ref{fig:controller}), a barrier
model (\ref{fig:barrier}) and four identical trains (\ref{fig:traini}). The resulting Petri Net is
obtained by the parallel composition of these \ttpn.
  
The property ``the barrier is closed when a train crosses the road'' is a safety property and is
interpreted as a reachability test: we want to check if there exists a state such that for any train
$i$: $M(On_i)=1$ and $M(Closed)=0$. This could be checked directly on the computed graph using
\mercutio or using \uppaal to test the property. In \uppaal, the property is expressed as:
\texttt{\small E<>((M[$On_1$]==1 or M[$On_2$]==1 or M[$On_3$]==1 or M[$On_4$]==1) and
  M[Closed]==0)}. In both cases, the result is \false, proving that no train may cross the road
while the barrier is not closed.

Using the automaton, it is possible to model time properties. For instance, ``when the train $i$
approaches, the barrier closes within delay $\delta$'' may be checked. In \tctl this property is
expressed by: $ M(close_i)= \; \uparrow 1 \implies \forall\Diamond_{\leq \delta} M(closed)=1$.
$M(close_i) = \; \uparrow 1$ means that only states for which $M(close_i)=1$ in the state and
$M(close_i)=0$ for all the preceding states. To check this property on the \ta using \uppaal or on
the \ttpn using reachability analysis leads to create an observer or modify the model. For instance,
to use \uppaal we have to add an additional clock that starts when a train change its state to
$close_i$. By using \kronos, there is no need to modify or create an observer. Given the \ta and a
\tctl formula, \kronos can perform model-checking using classical \tctl forward or backward
algorithms.

\subsection{Mixing Timed Automata and Time Petri Nets}

The method proposed in this paper provides a common framework for using and analyzing reactive
systems modeled with Timed Automata or Time Petri Nets.

Many systems are modeled using \ttpn (FIP, CAN), nevertheless some problems (time controller
synthesis for instance) benefit of larger studies and efficient tools. Then, it may be necessary to
have a mixed representation of the system.

We give here some examples of mixing Timed Automata and Time Petri Nets:

\begin{description}
\item[Test Case] Given a reactive system expressed with a \ttpn, different scenarios may be studied
  by synchronizing it with a Test Automaton. This Test Automaton represents the sequence of
  transitions to be fired and the synchronization is made over the firing of transitions.
\item[Controller] Given a reactive system expressed with a \ttpn, a controller may be modeled using
  \ta to constraint the execution of the system.
\end{description}

\section{Conclusions}
\label{sec:conclusions}
In this paper, we proposed an efficient method to compute the state space of a bounded \ttpn. The
proposed algorithm performs a forward computation of the state space and we proved it is exact with
respect to reachability even for bounded \ttpn with $\infty$ as latest firing time. We proposed a
labeling algorithm of the produced graph to build a Timed Automaton that we proved to be timed
bisimilar to the original \ttpn. Some examples were given to show that our tool performs better than
two other methods used to compute the state space of a \ttpn: the State Class Timed Automaton (\gpn)
and the State Class Graph (\tina). Though the number of clocks of our \ta is greater than the one of
the State Class Timed Automaton, our construction is faster and syntactical clocks reduction
algorithms may be successfully applied to reduce it.

Consequently, our method allows the use of Timed Automaton tools to model-check \ttpn. In
particular, the Timed Marking Automaton makes \tctl model-checking feasible for bounded \ttpn,
which, to our knowledge has not been done before.

We are currently involved in two different research area. First, we think possible to use efficient
data structures (BDD-like structure) to improve our implementation and we are studying Partial Order
methods to reduce time and space requirements. Finally, it would be useful to develop a full
model-checker for \ttpn without having to build the Timed Automaton. Then, a further step in the
analysis of real-time reactive systems will be to provide methods for the time controller synthesis
problem for \ttpn.

\bibliographystyle{acmtrans} \bibliography{gardey-thesis}

\begin{thebibliography}{}

\bibitem[\protect\citeauthoryear{Abdulla and Jonsson}{Abdulla and
  Jonsson}{1998}]{abdulla-TCS-01}
{\sc Abdulla, P.~A.} {\sc and} {\sc Jonsson, B.} 1998.
\newblock Ensuring completeness of symbolic verification methods for
  infinite-state systems.
\newblock {\em Theoretical Computer Science\/}~{\em 256}, 145--167.

\bibitem[\protect\citeauthoryear{Abdulla and Nylén}{Abdulla and
  Nylén}{2001}]{abdulla-ICATPN-01}
{\sc Abdulla, P.~A.} {\sc and} {\sc Nylén, A.} 2001.
\newblock Timed petri nets and bqos.
\newblock In {\em 22nd International Conference on Application and Theory of
  Petri Nets (ICATPN'01)}. Lecture Notes in Computer Science, vol. 2075.
  Springer-Verlag, Newcastle upon Tyne, United Kingdom, 53--70.

\bibitem[\protect\citeauthoryear{Alur and Dill}{Alur and
  Dill}{1994}]{alur-TCS-94}
{\sc Alur, R.} {\sc and} {\sc Dill, D.~L.} 1994.
\newblock A theory of timed automata.
\newblock {\em Theoretical Computer Science\/}~{\em 126,\/}~2, 183--235.

\bibitem[\protect\citeauthoryear{Berthomieu and Diaz}{Berthomieu and
  Diaz}{1991}]{berthomieu-IEEE-91}
{\sc Berthomieu, B.} {\sc and} {\sc Diaz, M.} 1991.
\newblock Modeling and verification of time dependent systems using time petri
  nets.
\newblock {\em IEEE transactions on software engineering\/}~{\em 17,\/}~3
  (March), 259--273.

\bibitem[\protect\citeauthoryear{Berthomieu and Vernadat}{Berthomieu and
  Vernadat}{2003}]{bertho-TACAS-03}
{\sc Berthomieu, B.} {\sc and} {\sc Vernadat, F.} 2003.
\newblock State class constructions for branching analysis of time petri nets.
\newblock In {\em 9th International Conference on Tools and Algorithms for the
  Construction and Analysis of Systems (TACAS 2003)}. Lecture Notes in Computer
  Science, vol. 2619. Springer Verlag, Warsaw, Poland, 442--457.

\bibitem[\protect\citeauthoryear{Bouyer}{Bouyer}{2002}]{bouyer-rr-02}
{\sc Bouyer, P.} 2002.
\newblock Timed automata may cause some troubles.
\newblock Tech. rep., LSV. July.

\bibitem[\protect\citeauthoryear{Bouyer}{Bouyer}{2003}]{bouyer-stacs-03}
{\sc Bouyer, P.} 2003.
\newblock Unteamable timed automata!
\newblock In {\em Proc. 20th Annual Symposium on Theoretical Aspects of
  Computer Science (STACS'2003)}. LNCS, vol. 2607. Springer Verlag, Berlin,
  Germany, 620--631.

\bibitem[\protect\citeauthoryear{Cassez and Roux}{Cassez and
  Roux}{2004}]{avocs-04}
{\sc Cassez, F.} {\sc and} {\sc Roux, O.~H.} 2004.
\newblock Structural translation from time {Petri} nets to timed automata.
\newblock In {\em Fourth International Workshop on Automated Verification of
  Critical Systems (AVoCS'04)}. London (UK).

\bibitem[\protect\citeauthoryear{Cortès, Eles, and Peng}{Cortès
  et~al\mbox{.}}{2000}]{cortes-isss-00}
{\sc Cortès, L.~A.}, {\sc Eles, P.}, {\sc and} {\sc Peng, Z.} 2000.
\newblock Verification of embedded systems using a petri net based
  representation.
\newblock In {\em 13th International Symposium on System Synthesis (ISSS
  2000)}. Madrid, Spain, 149--155.

\bibitem[\protect\citeauthoryear{Daws and Yovine}{Daws and
  Yovine}{1996}]{daws-yovine-96}
{\sc Daws, C.} {\sc and} {\sc Yovine, S.} 1996.
\newblock Reducing the number of clock variables of timed automata.
\newblock In {\em 17th IEEE Real Time Systems Symposium, RTSS'96}. IEEE
  Computer Society Press.

\bibitem[\protect\citeauthoryear{de~{Frutos Escrig}, Ruiz, and
  Alonso}{de~{Frutos Escrig} et~al\mbox{.}}{2000}]{defrutos-ICATPN-00}
{\sc de~{Frutos Escrig}, D.}, {\sc Ruiz, V.~V.}, {\sc and} {\sc Alonso, O.~M.}
  2000.
\newblock Decidability of properties of timed-arc petri nets.
\newblock In {\em 21st International Conference on Application and Theory of
  Petri Nets (ICATPN'00)}. Lecture Notes in Computer Science, vol. 1825.
  Springer-Verlag, Aarhus, Denmark, 187--206.

\bibitem[\protect\citeauthoryear{Diaz and Senac}{Diaz and
  Senac}{1994}]{Diaz-lncs-94}
{\sc Diaz, M.} {\sc and} {\sc Senac, P.} 1994.
\newblock Time stream petri nets: a model for timed multimedia information.
\newblock In {\em 15th International Conference on Application and Theroy of
  Petri Nets}. LNCS, vol. 815. Springer Verlag, Zaragoza, Spain, 219--238.

\bibitem[\protect\citeauthoryear{Finkel and Schnoebelen}{Finkel and
  Schnoebelen}{1998}]{finkel-LATIN-98}
{\sc Finkel, A.} {\sc and} {\sc Schnoebelen, P.} 1998.
\newblock Fundamental structures in well-structured infinite transitions
  systems.
\newblock In {\em 3rd Latin American Theoretical Informatics Symposium
  (LATIN'98)}. Lecture Notes in Computer Science, vol. 1380. Springer-Verlag,
  Campinas, Brazil, 102--118.

\bibitem[\protect\citeauthoryear{Gardey, Roux, and F.Roux}{Gardey
  et~al\mbox{.}}{2003}]{gardey-FORMATS-03}
{\sc Gardey, G.}, {\sc Roux, O.~H.}, {\sc and} {\sc F.Roux, O.} 2003.
\newblock Using zone graph method for computing the state space of a time petri
  net.
\newblock In {\em Formal Modeling and Analysis of Timed Systems
  (FORMATS'2003)}. LNCS. Springer--Verlag, Marseille, France.

\bibitem[\protect\citeauthoryear{Khansa, Denat, and Collart-Dutilleul}{Khansa
  et~al\mbox{.}}{1996}]{khansa-wodes-96}
{\sc Khansa, W.}, {\sc Denat, J.-P.}, {\sc and} {\sc Collart-Dutilleul, S.}
  1996.
\newblock P-time petri nets for manufacturing systems.
\newblock In {\em International Workshop on Discrete Event Systems, WODES'96}.
  Edinburgh (U.K.), 94--102.

\bibitem[\protect\citeauthoryear{Larsen, Pettersson, and Yi}{Larsen
  et~al\mbox{.}}{1997}]{uppaal}
{\sc Larsen, K.~G.}, {\sc Pettersson, P.}, {\sc and} {\sc Yi, W.} 1997.
\newblock {\sc Uppaal}\ in a nutshell.
\newblock {\em International Journal on Software Tools for Technology
  Transfer\/}~{\em 1,\/}~1--2 (Oct), 134--152.
\newblock http://www.uppaal.com/.

\bibitem[\protect\citeauthoryear{Lilius}{Lilius}{1999}]{lilius-ENTCS-99}
{\sc Lilius, J.} 1999.
\newblock Efficient state space search for time petri nets.
\newblock In {\em MFCS Workshop on Concurrency '98}. ENTCS, vol.~18. Elsevier.

\bibitem[\protect\citeauthoryear{Lime and Roux}{Lime and
  Roux}{2003}]{lime-roux-pnpm-03}
{\sc Lime, D.} {\sc and} {\sc Roux, O.~H.} 2003.
\newblock State class timed automaton of a time petri net.
\newblock In {\em The 10th International Workshop on Petri Nets and Performance
  Models, (PNPM'03)}. IEEE Computer Society.

\bibitem[\protect\citeauthoryear{Menasche}{Menasche}{1982}]{menasche-thesis-82}
{\sc Menasche, M.} 1982.
\newblock Analyse des réseaux de petri temporisés et application aux systèmes
  distribués.
\newblock Ph.D. thesis, Université Paul Sabatier, Toulouse, France.

\bibitem[\protect\citeauthoryear{Merlin}{Merlin}{1974}]{merlin-phd-74}
{\sc Merlin, P.~M.} 1974.
\newblock A study of the recoverability of computing systems.
\newblock Ph.D. thesis, Department of Information and Computer Science,
  University of California, Irvine, CA.

\bibitem[\protect\citeauthoryear{Okawa and Yoneda}{Okawa and
  Yoneda}{1997}]{okawa-yoneda-97}
{\sc Okawa, Y.} {\sc and} {\sc Yoneda, T.} 1997.
\newblock Symbolic ctl model checking of time petri nets.
\newblock In {\em Electronics and Communications in Japan}, {S.~Technica}, Ed.
  Vol.~80. 11--20.

\bibitem[\protect\citeauthoryear{Popova}{Popova}{1991}]{popova-JIPC-91}
{\sc Popova, L.} 1991.
\newblock On time petri nets.
\newblock {\em Journal Information Processing and Cybernetics, EIK\/}~{\em
  27,\/}~4, 227--244.

\bibitem[\protect\citeauthoryear{Ramchandani}{Ramchandani}{1974}]{ramchandani-%
phd-74}
{\sc Ramchandani, C.} 1974.
\newblock Analysis of asynchronous concurrent systems by timed {P}etri nets.
\newblock Ph.D. thesis, Massachusetts Institute of Technology, Cambridge, MA.
\newblock Project MAC Report MAC-TR-120.

\bibitem[\protect\citeauthoryear{Rokicki}{Rokicki}{1993}]{rokicki-phd}
{\sc Rokicki, T.~G.} 1993.
\newblock Representing an modeling circuits.
\newblock Ph.D. thesis, Stanford University.

\bibitem[\protect\citeauthoryear{Rokicki and Myers}{Rokicki and
  Myers}{1994}]{rokicki-cav-94}
{\sc Rokicki, T.~G.} {\sc and} {\sc Myers, C.~J.} 1994.
\newblock Automatic verification of timed circuits.
\newblock In {\em 6th International Conference on Computer-Aided Verification
  (CAV'94)}. LNCS, vol. 818. Springer--Verlag, 468--480.

\bibitem[\protect\citeauthoryear{Romeo}{Romeo}{2003}]{romeo}
{\sc Romeo}. 2003.
\newblock http://www.irccyn.ec-nantes.fr/irccyn/d/fr/equipes/tempsreel/logs.
\newblock {\em A tool for Time Petri Nets Analysis\/}.

\bibitem[\protect\citeauthoryear{Sava and Alla}{Sava and
  Alla}{2001}]{sava-alla-01}
{\sc Sava, A.~T.} {\sc and} {\sc Alla, H.} 2001.
\newblock Commande par supervision des systèmes à évènements discrets
  temporisées.
\newblock {\em Modélisation des systèmes réactifs (MSR 2001)\/}, 71--86.

\bibitem[\protect\citeauthoryear{Toussaint, Simonot-Lion, and
  Thomesse}{Toussaint et~al\mbox{.}}{1997}]{toussaint-FTDCS-97}
{\sc Toussaint, J.}, {\sc Simonot-Lion, F.}, {\sc and} {\sc Thomesse, J.-P.}
  1997.
\newblock Time constraint verifications methods based time petri nets.
\newblock In {\em 6th Workshop on Future Trends in Distributed Computing
  Systems (FTDCS'97)}. Tunis, Tunisia, 262--267.

\bibitem[\protect\citeauthoryear{Yoneda and Ryuba}{Yoneda and
  Ryuba}{1998}]{yoneda-IEICE-98}
{\sc Yoneda, T.} {\sc and} {\sc Ryuba, H.} 1998.
\newblock Ctl model checking of time petri nets using geometric regions.
\newblock {\em IEICE Transactions on Information and Systems\/}~{\em
  E99-D,\/}~3 (march), 297--396.

\bibitem[\protect\citeauthoryear{Yovine}{Yovine}{1997}]{kronos}
{\sc Yovine, S.} 1997.
\newblock Kronos: A verification tool for real-time systems.
\newblock {\em International Journal of Software Tools for Technology
  Transfer\/}~{\em 1,\/}~1--2 (Oct), 123--133.

\end{thebibliography}

\end{document}